\documentstyle[12pt]{article}
\begin{document}

\begin{flushright}
June 2000

OU-HET 351
\end{flushright}

\begin{center}

\vspace{5cm}
{\Large Description of Intersecting Branes} 

\vspace{3mm}

{\Large via Tachyon Condensation}

\vspace{2cm}
Takao Suyama \footnote{e-mail address : suyama@funpth.phys.sci.osaka-u.ac.jp}

\vspace{1cm}

{\it Department of Physics, Graduate School of Science, Osaka University, }

{\it Toyonaka, Osaka, 560-0043, Japan}

\vspace{4cm}

{\bf Abstract} 

\end{center}

We construct a model describing BPS brane-systems using low energy effective theory of 
brane-antibrane system. 
Both parallel branes and intersecting branes can be treated by this model. 
After tachyon condensation, the dynamics of fluctuations around such brane-systems is 
supersymmetric if the degrees of freedom are restricted on the branes. 
The form of the tachyon potential and the application of this model to the black hole 
physics are discussed. 

\newpage

{\bf {\large 1. Introduction}}

\vspace{5mm}

Low energy dynamics of the combined D-brane systems is important, for example, for  
black hole physics \cite{BlackHole}. 
It is easy to describe such system which contains only one kind of D-branes. 
The effective theory of the system is the dimensional reduction of the ten-dimensional 
super Yang-Mills theory \cite{braneLEET}. 
However, construction of effective theory is not so straightforward when the system 
consists of a combination of D-branes with different dimensions. 
In most cases, one pays attention to a specific part of the worldvolume and consider the 
effective theory living only on that region \cite{BlackHole}\cite{suyama}. 
For example, the effective theory of the D1-D5 system is a two-dimensional superconformal 
field theory living on the D1-branes. 
Although this is enough to discuss the physics of the black holes, it is preferable to have 
another way of describing the system as a whole. 

The difficulty in such description may be how to treat the brane-intersections. 
When one writes down the effective theory of the massless modes on the whole brane-system, 
the theory contains a delta-function-like interaction at the intersection.  
It would not be an easy task to deal with such theory, although this description has 
some applications \cite{deltaint}. 

In this paper we propose another way of describing the D-brane system which includes
intersecting branes. 
Our idea is to apply the tachyon condensation \cite{nonBPS}. 
In this situation, pair annihilations of D-branes can be discussed and in some cases 
the lower dimensional branes can remain as a result.
The worldvolume of the resulting brane is determined by the form of the tachyon field 
and, in particular, it need not be a smooth manifold. 
Therefore the theory describing the above phenomenon would be able to handle the low energy 
dynamics of the intersecting branes. 

This paper is organized as follows. 
We review the story of the tachyon condensation briefly in section 2. 
The construction of the model is discussed in section 3. 
This model can describe both parallel branes (section 4) and intersecting branes 
(section 5). 
Section 6 is devoted to discussions. 
In Appendix, we summarize the properties of the vortex solutions used in this paper. 

\vspace{1cm}

{\bf \large {2. Review of the tachyon condensation}}

\vspace{5mm}

In this section, we will review some properties of Dp-$\bar{\mbox{D}}$p system. 
While both branes preserve half of the supersymmetries,  
this system breaks all of the supersymmetries when they exist together. 

There exists a tachyonic mode coming from the string stretched between Dp-brane and 
$\bar{\mbox{D}}$p-brane, which survives the GSO projection \cite{tachyon}. 
This indicates the instability of the system. 
Assuming that the tachyon potential has a minimum, the tachyon would condense and 
this system decays to some stable vacuum. 

Let $T$ denote the tachyon field whose potential $V(T)$ has a minimum at $T=T_0$. 
When $T=T_0$ everywhere, there remains nothing : pair annihilation of the D-branes occurs. 
This has been checked by examining that $V(T_0)$ would exactly cancel the tension of the 
D-branes, using string field theory \cite{SFT}. 
On the other hand, when $T$ has some nontrivial form, there remains a lower dimensional 
brane which is stable but not necessarily BPS \cite{nonBPS}. 
In this case, the tension of the originally unstable brane-antibrane system is cancelled 
in almost all regions except at those places where the tachyon field behaves nontrivially. 
The energy therefore remains nonzero only on that regions, and this is regarded as a 
lower dimensional brane. 
It is argued that K-theory determines what kind of branes can appear after tachyon 
condensation \cite{Witten}\cite{Horava}. 

For example, consider the D1-$\bar{\mbox{D}}$1 system extended along $x^1$ direction 
\cite{nonBPS}. 
Suppose that $T$ supports a kink,
$$ T(x)= \left\{ 
\begin{array}{l} 
  +T_0 \hspace{5mm} (x^1\to +\infty) \\ -T_0 \hspace{5mm} (x^1\to -\infty)
\end{array} \right. $$
$$ T(x^1=0)=0,$$
where we have assumed that $V(-T)=V(T)$ and $T=-T_0$ is also a minimum of $V(T)$. 
Then the remaining energy is localized around $x^1=0$, which can be regarded as D0-brane. 
In ordinary Type IIB theory this is of course unstable, but this is stable in 
Type I theory \cite{nonBPS}.

The brane annihilation summarized above is also described in terms of gauge theory on the 
D-brane worldvolume \cite{Horava}. 
The worldvolume theory would be a gauge theory coupled to the tachyon. 
When the tachyon acquires VEV, the gauge symmetry is broken down by the Higgs mechanism. 
Thus the gauge fields acquire masses and are decoupled from the low energy effective theory.   
This indicates the disappearance of the D-branes. 
The case of nontrivial $T$ can be described similarly. 
If $T=0$ in some region, the gauge symmetry is recovered there, and this corresponds to the 
resulting lower dimensional brane. 

In the gauge theory language, the classification of the remaining branes can be done in 
terms of homotopy groups of the vacuum manifold \cite{Horava}. 

\vspace{1cm}

{\bf {\large 3. The model}}

\vspace{5mm}

We will consider in this section the model of D9-$\bar{\mbox{D}}$9 system in 
Type IIB theory. 
The model of Dp-$\bar{\mbox{D}}$p system can be obtained by dimensional reduction. 

There is a $U(1)$ gauge multiplet on each brane and also a tachyon field $T$ which couples to 
both gauge fields. 
One linear combination of these gauge fields (so called diagonal $U(1)$) will decouple from 
the dynamics (it is argued that this gauge field becomes a fundamental string 
\cite{diagonalU(1)}, and its generalizations are discussed \cite{general}).
Then the resulting model is ten-dimensional $U(1)$ gauge theory 
coupled to the tachyon field,
\begin{equation}
S = \int d^{10}x \left\{ -\frac14 F_{\mu\nu}F^{\mu\nu} +\frac i2 \bar{\psi}\Gamma^{\mu}
\partial_{\mu} \psi -|D_{\mu}T|^2 - V(T) \right\} \label{action}
\end{equation}
where $D_{\mu}T=\partial_{\mu}T+iA_{\mu}T$. 
$\Gamma^{\mu}$ are the ten-dimensional gamma matrices and $\psi$ is a ten-dimensional 
Majorana-Weyl spinor.
Although there exists a charged massless fermion coming from the string stretched between 
branes, it is not important for the following discussions and we will ignore it. 
This action would be a first few terms in the expansion of the more complicated DBI-like 
action. 
The spacetime action of non-BPS brane is argued in \cite{nonBPSaction}. 
It is determined by requiring that it reduces to the ordinary DBI action if 
the tachyon and half of the massless fermions which are absent on the BPS brane are set 
to zero. 
This seems to imply that in the Dp-$\bar{\mbox{D}}$p case the action is also 
supersymmetric if the tachyon field $T$ is set to zero. 
The action of this system is discussed in \cite{action2}. 

The action (\ref{action}) is obviously not supersymmetric. 
According to the discussion in string theory, however, classical solutions should exist 
which preserve some of the supersymmetries. 
The first two terms are invariant under the ordinary transformations.
\begin{eqnarray}
&& \delta A_{\mu} = \frac i2 \bar{\epsilon}\Gamma_{\mu}\psi \label{transfA} \\
&& \delta \psi\ = -\frac14 F_{\mu\nu}\Gamma^{\mu\nu}\epsilon \label{transf}
\end{eqnarray}
The tachyon should be invariant : $\delta T = 0$. Then,
\begin{equation}
\delta S = \int d^{10}x \frac i2 \bar{\epsilon}\Gamma_{\mu}\psi (iT^{\dag}D^{\mu}T-
             iD^{\mu}T^{\dag}T).
\end{equation}
This vanishes when $D_{\mu}T=0$ which leads to trivial solutions. 

Now we will modify the transformations of the fermions (\ref{transf}) as follows,
\begin{equation}
\delta \psi = -\frac14 F_{\mu\nu}\Gamma^{\mu\nu}\epsilon + f(|T|^2)\epsilon, \label{transfF}
\end{equation}
where $f(x)$ is a function which may have spinor indices. 
Then the variation of the action is 
\begin{equation}
\delta S = \int d^{10}x \frac i2 \bar{\psi}\Gamma_{\mu} (-iT^{\dag}D^{\mu}T+
             iD^{\mu}T^{\dag}T + 2\partial_{\mu}f(|T|^2))\epsilon.
\end{equation}
We will show below that for a suitable choice of $f(x)$, there exist classical 
solutions which preserve some of supersymmetries. 

\vspace{1cm}

{\bf {\large 4. Parallel D-branes}}

\vspace{5mm}

We will consider for example D4-$\bar{\mbox{D}}$4 system which decays to D2-branes. 
Suppose that D4-brane and $\bar{\mbox{D}}$4-brane extend along the (01234) directions, 
and the resulting D2-branes extend along (034) directions. 
The action of the model is eq.(\ref{action}) dimensionally reduced to five dimensions. 
For the solutions corresponding to such branes to exist, the appropriate choice of $f(x)$ is 
\begin{equation}
f(|T|^2) = -\frac12 s(|T|^2-\zeta)\Gamma^1\Gamma^2 \hspace{5mm} (\zeta >0) .
\label{mdfiedtransf}
\end{equation}
The variation of the action is 
\begin{eqnarray}
&& \delta S = -\int d^5x \ \Bigl[ \frac i2 \bar{\psi}T^{\dag} \Bigl( \Gamma^1(iD_1T+sD_2T)
              +\Gamma^2(iD_2T-sD_1T) \\ \nonumber
&& \hspace*{2cm} +D_iT(i\Gamma^i+\Gamma^{i12})+iA_kT(i\Gamma^k+\Gamma^{k12}) 
               \Bigr) \epsilon + (h.c.) \Bigr],
\end{eqnarray}
where $i=0,3,4$ and $k=5,6,\cdots,9$. 
In this case $A_k$ is the neutral scalars. 
Thus the conditions for $\delta S=0$ are
\begin{eqnarray}
&& D_1T-isD_2T = 0 \label{condition}\\
&& \hspace{8mm}D_iT = 0 \label{eq1} \\
&& \hspace{8mm}A_kT = 0 \label{eq2},
\end{eqnarray}
We assume translational invariance along $i$-th direction, thus eq.(\ref{eq1}) sets 
$A_i=0$, and also $A_k=0$ by eq.(\ref{eq2}). 
$\delta \psi=0$ means 
\begin{equation}
F_{12}+s(|T|^2-\zeta) = 0 .
\end{equation}
We will investigate the stability of this BPS solutions. 
Their energy is rewritten as follows.
\begin{equation}
E = \int d^4x \left\{\frac12 \left( F_{12}+s(|T|^2-\zeta)\right)^2+|D_1T-isD_2T|^2 
      +s\zeta F_{12} +V(T) -\frac12(|T|^2-\zeta)^2 \right\} \label{energy}
\end{equation}
If the tachyon potential takes the form
\begin{equation}
V(T) = \frac12 (|T|^2-\zeta)^2, \label{tachyonV}
\end{equation}
then eq.(\ref{energy}) becomes
\begin{equation}
E = \int d^4x \left\{ \frac12 \left(F_{12}+s(|T|^2-\zeta)\right)^2+|D_1T-isD_2T|^2 
+ s\zeta F_{12} \right\}.
\end{equation}
Therefore the BPS solutions are stable topologically. 

Now we have the following BPS equations whose solutions preserve 16 supercharges.
\begin{eqnarray}
&& D_1T-isD_2T = 0 \\
&& F_{12}+s(|T|^2-\zeta) = 0
\end{eqnarray}
This is the equations for the Nielsen-Olesen vortex \cite{vortex}. 
The properties of this solutions are well-known. 
Some of these are collected in the Appendix. 
The solutions are labeled by an integer $n$ (quantized magnetic flux), and is determined 
by specifying $n$ distinct points in 1-2 plane, at which $T=0$. 
As explained in section 2, zero loci of the tachyon field $T$ correspond to the D-brane 
worldvolume. 
Thus the $n$-vortex solution describes $n$ parallel D2-branes. 

The worldvolume theory on the D2-branes should be a supersymmetric one. 
For this, eqs.(\ref{condition})(\ref{eq1})(\ref{eq2}) has to be satisfied even when the 
fluctuations are included. 
Let $a_{\mu},\varphi$ denote the fluctuations of $A_{\mu},\psi$ around the vortex solution 
respectively, then eqs.(\ref{condition})(\ref{eq1})(\ref{eq2}) means
\begin{equation}
a_{\mu} \ne 0  \hspace{5mm}\mbox{only at} \hspace{5mm} T=0
\end{equation}
From the transformations (\ref{transfA}), $\varphi$ is also restricted to the region 
specified by $T=0$. 
Then the resulting model is supersymmetric and the physical degrees of freedom exist only 
at the cores of the vortices, as is expected from the D-brane interpretation. 

It is interesting that we can construct $n$ D-brane system from $U(1)$ gauge theory. 
Moreover system with diferent number of D-branes merely corresponds to taking another 
classical solution in the same model. 
In the ordinary approach to the D-brane worldvolume theory, the number of D-branes must be 
fixed to construct the theory. 
It might be possible to relate this feature of our model to the second 
quantization of D-branes. 

\vspace{1cm}

{\bf {\large 5. Intersecting branes}}

\vspace{5mm}

We will show in this section that our model can also describe intersecting branes 
in a similar way which was discussed in the previous section. 
We will consider the D4-$\bar{\mbox{D}}$4 system which extends along (01234) directions. 
The action of the model is again eq.(\ref{action}) dimensionally reduced to five dimensions, 
and the tachyon potential and supersymmetry transformations are given as 
eqs.(\ref{tachyonV})(\ref{transfF})(\ref{mdfiedtransf}). 
Suppose that this system decays to the intersecting D2-D2' system, in which D2-branes 
extend alnog (034) directions and D2'-branes along (012) directions. 
The conditions for the remaining supersymmetries are 
\begin{eqnarray}
&& \Gamma^1\Gamma^2\epsilon = s'\Gamma^3\Gamma^4\epsilon \\
&& (s' = \pm 1). \nonumber
\end{eqnarray}
The BPS conditions are then
\begin{eqnarray}
&& D_1T-isD_2T = 0 \label{tachyoneq1} \\
&& D_3T-iss'D_4T = 0 \label{tachyoneq2} \\
&& F_{12}+s'F_{34}+s(|T|^2-\zeta) = 0 \label{nontrivial} \\
&& F_{13}-s'F_{24} = 0 \label{redundant1}\\
&& F_{14}+s'F_{23} = 0, \label{redundant2}
\end{eqnarray}
where we have assumed $A_k=0 \hspace{2mm}(k=0,5,6,\cdots,9)$ and that the solutions are 
static. 
It is easy to check that the solutions of these equations are stable topologically. 

At first sight, these conditions are overdetermined. 
However the last two equations (\ref{redundant1})(\ref{redundant2}) are in fact redundant. 
From eqs.(\ref{tachyoneq1})(\ref{tachyoneq2}), $T$ and $A_i \hspace{2mm}(i=1,2,3,4)$ are 
written as follows.
\begin{eqnarray}
&& T = \rho^{\frac12} e^{i\omega} \\
&& A_1 = \ \ \frac12 s\partial_2\log\rho -\partial_1\omega \\
&& A_2 = -\frac12 s\partial_1\log\rho-\partial_2\omega \\
&& A_3 = \ \ \frac12 ss'\partial_4\log\rho-\partial_3\omega \\
&& A_4 = -\frac12 ss'\partial_3\log\rho-\partial_4\omega,
\end{eqnarray}
These solve eqs.(\ref{redundant1})(\ref{redundant2}) automatically. 
The only nontrivial equation (\ref{nontrivial}) is then
\begin{equation}
-\frac12 \sum_{i=1}^4 (\partial_i)^2 \log\rho + \rho - \zeta = 0. \label{rewrite}
\end{equation}
$\omega$ is determind by requiring the regularity of $A_i$ and the single-valuedness of 
$T$. 

The solutions we would like to find are symmetric under the rotations in 1-2 and 3-4 
planes (D2(D2')-branes coincide).  
Now we take the polar coordinates for these planes,
$$ (x_1,x_2) \to (r_1,\theta_1)\ ,\  (x_3,x_4) \to (r_2,\theta_2),$$
and assume that $\rho$ is independent of both $\theta_1$ and $\theta_2$. 
Then (\ref{rewrite}) becomes 
\begin{equation}
\left[ \frac{\partial^2}{\partial r_1^2}+\frac1{r_1}\frac{\partial}{\partial r_1} 
+\frac{\partial^2}{\partial r_2^2}+\frac1{r_2}\frac{\partial}{\partial r_2} \right] \log\rho 
= 2(\rho - \zeta).
\end{equation}
The region around $r_1=r_2=0$, $\rho$ behaves as
\begin{equation}
\rho \sim r_1^{2n_1}r_2^{2n_2}. \label{origin}
\end{equation}
For the regularity and the single-valuedness of the solution, $n_1$ and $n_2$ are positive 
integers. 

It is convenient to make the following change of variables : $r_1=e^{t_1},r_2=e^{t_2}$. 
Then
\begin{equation}
\left[ e^{-2t_1}\frac{\partial^2}{\partial t_1^2}+e^{-2t_2}\frac{\partial^2}{\partial t_2^2}
\right] \log\rho = 2(\rho-\zeta). \label{modified}
\end{equation}
When $t_2 \to \infty$, the second term of LHS is negligible and $\rho$ becomes $t_2$ (thus 
$r_2$) independent. 
This means that in this region $\rho$ approaches the vortex solution with magnetic flux 
$n_1$ discussed in the previous section. 
The same is true with the behavior of $\rho$ when $t_1\to \infty \ (r_1\to \infty)$ and 
its flux is $n_2$. 
By the continuity of $\omega$, $n_1$ and $n_2$ take the same values as the ones in 
eq.(\ref{origin}). 

The global behavior of the solutions can also be discussed. 
Now take $u=\log\frac{\rho}{\zeta}$ and define ${\cal D}_+$ to be the region where $u>0$. 
${\cal D}_+$ is a finite region in ${\bf R}^2$. 
Then using eq.(\ref{modified}),
\begin{eqnarray}
0 &=& \int_{{\cal D}_+} dt_1dt_2 \left[ u\{-e^{2t_2}\partial_1^2u-e^{2t_1}\partial_2^2
       +2\zeta e^{2t_1+2t_2}(e^u-1)\} \right] \nonumber \\
  &=& \int_{{\cal D}_+} dt_1dt_2 \left[ e^{2t_2}(\partial_1u)^2+e^{2t_1}(\partial_2u)^2
       +2\zeta e^{2t_1+2t_2}u(e^u-1) \right] , \label{positivity}
\end{eqnarray}
where we have used that $u=0$ at the boundary of ${\cal D}_+$. 
The integrand is strictly positive, and therefore, eq.(\ref{positivity}) means $u\le 0$ 
(i.e. $\rho \le \zeta$) everywhere. 

The above analyses imply that $\rho=0$ at $r_1=0$ and/or $r_2=0$, and $\rho\sim \zeta$ away 
from the cores of the vortices. 
$\omega$ is determined to be $n_1\theta_1+n_2\theta_2$, and this gives
\begin{eqnarray}
\frac1{2\pi} \int dx_1dx_2 F_{12} &=& n_1 \\
\frac1{2\pi} \int dx_3dx_4 F_{34} &=& n_2.
\end{eqnarray}
Therefore this solution corresponds to $n_1$ coincident D2-branes and $n_2$ coincident D2'-
branes. 

The dynamics of the fluctuations around this solution is supersymmetric if they are 
restricted at the cores of the vortices, as in the case of parallel D-branes. 
This means that we can, in principle, construct the effective theory of D-brane system 
whose worldvolume is not a smooth manifold. 

\vspace{5mm}

The intersecting D4-D4'-D4'' system can also be described, starting from 
D6-$\bar{\mbox{D}}$6 system. 
The BPS conditions for this case are
\begin{eqnarray}
&& D_1T-isD_2T = 0 \nonumber \\
&& D_3T-iss'D_4T = 0 \nonumber \\
&& D_5T-iss''D_6T = 0 \nonumber \\
&& F_{12}+s'F_{34}+s''F_{56}+s(|T|^2-\zeta) = 0 \nonumber \\
&& F_{13}-s'F_{24} = 0 \nonumber \\
&& F_{14}+s'F_{23} = 0 \nonumber \\
&& F_{15}-s''F_{26} = 0 \nonumber \\
&& F_{16}+s''F_{25} = 0 \nonumber \\
&& s'F_{35}-s''F_{46} = 0 \nonumber \\
&& s'F_{36}+s''F_{45} = 0 \nonumber \\
&& \  (s',s''=\pm1).\nonumber
\end{eqnarray}
As in the D2-D2' case, the last six equations are redundant and the only nontrivial equation 
is rewritten as follows,
$$ -\frac12 \sum_{k=1}^6 (\partial_k)^2\log\rho+\rho-\zeta = 0, $$
where $T=\rho^{\frac12} e^{i\omega}$. 
The analysis of the solutions can be done similarly as for the previous case. 

\vspace{1cm}

{\bf {\large 6. Discussions}}

\vspace{5mm}

We have discussed the discription of the BPS brane-system via the tachyon condensation. 
The model (\ref{action}) can describe both parallel branes and intersecting branes. 
Moreover the system with different number of D-branes can be treated in the same model. 
The dynamics of the fluctuations which localize on the branes is supersymmetric, and 
this model can, in principle, provide a way of describing the effective theory of the 
brane-system whose worldvolume is not a smooth manifold. 

For the BPS D-branes to exist, we have seen that the specific form of the tachyon potential 
is needed. 
The existence of BPS D-branes would also require similar restriction on the tachyon potential 
in the case of more complicated DBI-like action \cite{nonBPSaction}. 
This might provide some information of the profile of the tachyon potential. 

D4-D4'-D4'' system is considered in the end of the previous section. 
To this system, one can add D0-branes without breaking any supersymmetries. 
The bound states of such a system was discussed in \cite{N=2} and it is conjectured that 
this has four bosonic states and four fermionic states (although the system considered in 
\cite{N=2} is the D-branes wrapped on some cycle in a Calabi-Yau manifold). 
If the model (\ref{action}) is generalized to the non-Abelian gauge thoery, there would 
exist a solution which contains the D0-branes in addition to the D4-branes. 
Therefore, the above conjecture could be checked by quantizing the collective coordinates 
of such a solution. 

As shown in section 3, $n$ parallel D-branes are described in terms of $U(1)$ gauge theory. 
The generic vortex solution corresponds to the separated D-branes and the gauge 
symmetry on each worldvolume is $U(1)$. 
It should be expected that when two or more vortices coincide, the gauge symmetry is 
enhanced. 
This will certainly be achieved in the non-Abelian model. 
This gauge enhancement might occur already in the $U(1)$ model, however. 

\vspace{2cm}

{\bf {\large Acknowledgements}}

\vspace{5mm}

I would like to thank H.Itoyama, T.Matsuo, K.Murakami for valuable discussions.
This work is supported in part by JSPS Research Fellowships.

\newpage

{\bf {\large Appendix : Vortex solutions}}

\vspace{5mm}

The Nielsen-Olesen vortex solution is the solution of the following equations 
\cite{vortex},
\begin{eqnarray}
&& D_1T-iD_2T = 0 \\
&& F_{12} + |T|^2-\zeta = 0 ,
\end{eqnarray}
where $D_kT=\partial_kT+iA_kT \ (k=1,2)$. 
$A_k$ and $T$ can be written as 
\begin{eqnarray}
&& T = \rho^{\frac12} e^{i\omega} \\
&& A_k = \frac12\varepsilon_{kl}\partial_l\log\rho-\partial_k\omega.
\end{eqnarray}
$\rho$ is determined by
\begin{equation}
-\frac12(\partial_1^2+\partial_2^2)\log\rho + \rho-\zeta = 0 \label{vortexeq}.
\end{equation}
We now consider the radially symmetric solutions. 
Taking the polar coordinate $(r,\theta)$ and assuming that $\rho$ is independent of $\theta$, 
(\ref{vortexeq}) becomes
\begin{equation}
-\frac12 \left( \frac{d^2}{dr^2}+\frac1r\frac d{dr} \right) \log\rho + \rho-\zeta =0.
\end{equation}
$\rho$ is a monotonic function and behaves as

$$ \rho \sim \left\{
  \begin{array}{l}
     r^{n} \hspace{5mm}(r\to 0) \\ \zeta \hspace{7mm}(r\to \infty)
  \end{array} \right. ,
$$
where $n$ is a positive integer.
This behavior makes $A_k$ singular at $r=0$. 
To eliminate this singularity, 
\begin{equation}
\omega=\frac n2\theta .
\end{equation} 
For the single-valuedness of $T$, $n$ must be an even integer. 

General solutions are also known \cite{vortex}. 
The $n$-vortex solution is determined by specifying the positions of $n$ points in the 
plane, at which $T=0$ (cores of the vortices).


\newpage

\begin{thebibliography}{99}

\bibitem{BlackHole}A.Strominger, C.Vafa, {\it Microscopic Origin of Bekenstein-Hawking 
Entropy}, Phys. Lett. {\bf B379} (1996) 99, hep-th/9601029; \\
C.Callan, J.Maldacena, {\it D-brane Approach to Black Hole Quantum Mechanics}, 
Nucl. Phys. {\bf B472} (1996) 591, hep-th/9601029.

\bibitem{braneLEET}E.Witten, {\it Bound States of Strings and p-Branes},Nucl. Phys. 
{\bf B460} (1996) 335, hep-th/9510135.

\bibitem{suyama}T.Suyama, {\it Monopoles and Black Hole Entropy}, Mod. Phys. Lett. {\bf A15} 
(2000) 271, hep-th/9909091.

\bibitem{deltaint}A.Kapustin, S.Sethi, {\it The Higgs Branch of Impurity Theories}, 
Adv. Theor. Math. Phys. {\bf 2} (1998) 571, hep-th/9804027.

\bibitem{tachyon}M.Green, {\it POINT-LIKE STATES FOR TYPE 2b SUPERSTRINGS}, Phys. Lett. 
{\bf B329} (1994) 435, hep-th/9403040; \\
T.Banks, L.Susskind, {\it Brane - Anti-Brane Forces}, hep-th/9511194; \\
M.Green, M.Gutperle, {\it Light-cone supersymmetry and D-branes}, Nucl. Phys. 
{\bf B476} (1996) 484, hep-th/9604091; \\
G.Lifschytz, {\it Comparing D-branes to Black-branes}, Phys. Lett. {\bf B388} (1996) 720, 
hep-th/9604156; \\
V.Periwal, {\it Antibranes and crossing symmetry}, hep-th/9612215.

\bibitem{SFT} A.Sen, B.Zwiebach, {\it Tachyon condensation in string field theory}, 
JHEP 0003 (2000) 002, hep-th/9912249 ; \\
W.Taylor, {\it D-brane effective field theory from string field theory}, hep-th/0001201; \\
N.Moeller, W.Taylor, {\it Level truncation and the tachyon in open bosonic string field 
theory}, hep-th/0002237; \\
J.Harvey, P.Kraus, {\it D-Branes as Unstable Lumps in Bosonic Open String Field Theory}, 
JHEP 0004 (2000) 012,hep-th/0002117; \\
R.de Mello Koch, A.Jevicki, M.Mihailescu, R.Tatar, {\it Lumps and P-branes in Open String 
Field Theory}, hep-th/0003031; \\
N.Berkovits, {\it The Tachyon Potential in Open Neveu-Schwarz String Field Theory}, 
JHEP 0004 (2000) 022, hep-th/0001084; \\
N.Berkovits, A.Sen, B.Zwiebach, {\it Tachyon Condensation in Superstring Field Theory}, 
hep-th/0002211; \\
P.De Smet, J.Raeymaekers, {\it Level Four Approximation to the Tachyon Potential in 
Superstring Field Theory}, hep-th/0003220; \\
N.Moeller, A.Sen, B.Zwiebach, {\it D-branes as Tachyon Lumps in String Field Theory}, 
hep-th/0005036; \\
A.Iqbal, A.Naqvi, {\it Tachyon Condensation on a non-BPS D-brane}, 
hep-th/0004015.

\bibitem{nonBPS}A.Sen, {\it Tachyon Condensation on the Brane Antibrane System}, 
JHEP 9808 (1998) 012, hep-th/9805170; \\
A.Sen, {\it SO(32) Spinors of Type I and Other Solitons on Brane-Antibrane Pair}, 
JHEP 9809 (1998) 023, hep-th/9808141. 

\bibitem{Witten}E.Witten, {\it D-Branes And K-Theory}, JHEP 9812 (1998) 019, 
hep-th/9810188.

\bibitem{Horava}P.Horava, {\it Type IIA D-Branes, K-Theory, and Matrix Theory}, Adv. 
Theor. Math. Phys. {\bf 2} (1999) 1373, hep-th/9812135.

\bibitem{diagonalU(1)}P.Yi, {\it Membranes from Five-Branes and Fundamental Strings from Dp
Branes},  Nucl.Phys. {\bf B550} (1999) 214, hep-th/9901159; \\
O.Bergman, K.Hori, P.Yi, {\it Confinement on the Brane}, hep-th/0002223.

\bibitem{general}L.Houart, Y.Lozano, {\it Type II Branes from Brane-Antibrane in M-theory}, 
Nucl.Phys. {\bf B575} (2000) 195, hep-th/9910266. 

\bibitem{nonBPSaction}A.Sen, {\it Supersymmetric World-volume Action for Non-BPS D-branes}, 
JHEP 9910 (1999) 008, hep-th/9909062; \\
M.Garousi, {\it Tachyon couplings on non-BPS D-branes and Dirac-Born-Infeld action}, 
hep-th/0003122; \\
E.Bergshoeff, M.de Roo, T.de Wit, E.Eyras, S.Panda, {\it T-duality and Actions for 
Non-BPS D-branes}, JHEP 0005 (2000) 009, hep-th/0003221.

\bibitem{action2}J.Kluson, {\it D-Branes in Type IIA and Type IIB Theories from Tachyon
Condensation}, hep-th/0001123.

\bibitem{vortex}H.Nielsen, P.Olesen, {\it Vortex Line Models for Dual Strings}, 
Nucl. Phys. {\bf B61} (1973) 45; \\
H.de Vega, F.Schaposnik, {\it Classical Vortex Solution of the Abelian Higgs Model}, 
Phys. Rev. {\bf D14} (1976) 1100; \\
E.Weinberg, {\it Multivortex Solutions of the Ginzburg-Landau Equations}, 
Phys. Rev. {\bf D19} (1979) 3008; \\
C.Taubes, {\it Arbitrary N-Vortex Solutions to the First Order Ginzburg-Landau Equations}, 
Comm. Math. Phys. {\bf 72} (1980) 277.

\bibitem{N=2}J.Maldacena, {\it N=2 Extremal Black Holes and Intersecting Branes}, 
Phys. Lett. {\bf B403} (1997) 20, hep-th/9611163.

\end{thebibliography}
\end{document}